\documentclass[preprint,showpacs,preprintnumbers,amsmath,amssymb]{revtex4}

\usepackage{graphicx}
\usepackage{dcolumn}
\usepackage{bm}

\usepackage{amssymb,bm}

\def\eqref#1{Eq.~(\ref{#1})}

\def\Eq#1{\begin{equation} #1 \end{equation}}
\def\Eqr#1{\begin{eqnarray} #1 \end{eqnarray}}

\newcommand{\nn}{\nonumber}
\newcommand{\pd}{\partial}

\def\Xsp{{\rm X}}

\def\Ysp{{\rm Y}}
\def\Zsp{{\rm Z}}
\def\X5sp{{\rm X}_5}
\def\Y3sp{{\rm Y}_3}
\def\Z3sp{{\rm Z}_3}

\def\lap{{\triangle}}

\begin{document}

\preprint{YITP-07-86}

\title{Dynamical D4-D8 and D3-D7 branes in supergravity}

\author{Pierre Binetruy}%

\affiliation{%
Astroparticule et Cosmologie, 
Universit\'e Paris Diderot, CNRS/IN2P3, CEA/DSM, Observatoire de Paris, 
B\^atiment Condorcet 
10, rue Alice Domon et L\'eonie Duquet, 75205 Paris Cedex 13, France.
}%

\author{Misao Sasaki}

\affiliation{
Yukawa Institute for Theoretical Physics
Kyoto University, Kyoto 606-8502, Japan.
}%

\author{Kunihito Uzawa}
 \altaffiliation[Also at ]{Yukawa Institute for Theoretical Physics
Kyoto University, Kyoto 606-8502, Japan.}

\affiliation{%
Osaka City University Advanced Mathematical Institute, 
Osaka 558-8585, Japan.
}%

\date{\today}

\begin{abstract}
We present a class of dynamical solutions for intersecting 
D4-D8 and D3-D7 brane systems in ten-dimensional type IIA and IIB 
supergravity. 
We discuss if these solutions can be recovered in
lower-dimensional effective theories for the warped compactification
of a general p-brane system.
It is found that an effective $p+1$-dimensional description
is not possible in general due to the entanglement
of the transverse coordinates and the $p+1$-dimensional 
coordinates in the metric components.
For the D4-D8 brane system, the dynamical solutions reduces to a 
static warped ${\rm AdS_6}\times {\rm S}^4$ geometry in a 
certain spacetime region. For the D3-D7 brane system,
we find a dynamical solution whose metric form is similar to 
that of a D3-brane solution. The main difference is
the existence of a nontrivial dilaton configuration
 in the D3-D7 solution.
Then we discuss cosmology of these solutions. We find 
that they behave like a Kasner-type cosmological
solution at $\tau\to\infty$, while it reduces to a warped static 
solution at $\tau\to0$, where $\tau$ is the cosmic time.
\end{abstract}

\pacs{11.25.-w, 11.27.+d, 98.80.Cq}
\maketitle


\section{Introduction}
 \label{sec:introduction}
Recently, studies on dynamical solutions of supergravity have been
a topic of great interest.
Conventionally time-dependent solutions of higher-dimensional 
supergravity are discussed in the context of lower-dimensional
effective theories after compactifying the internal space.
However, it is unclear how far this effective low-dimensional
description is valid. Thus, it is much more desirable to discuss
the four-dimensional cosmology in terms of the dynamics of the
original higher-dimensional theory. This is particularly true in
string cosmology in which the behavior of the early Universe is to be
understood in the light of string theory. Indeed, it was pointed out
that the four-dimensional effective theory for warped 
compactification of ten-dimensional type IIB supergravity 
allows solutions that cannot be obtained from solutions in 
the original higher-dimensional theories~\cite{Kodama:2005cz}. 

A time-dependent solution with fiveform flux 
 in the ten-dimensional type IIB supergravity was 
obtained by Gibbons, Lu and Pope~\cite{Gibbons:2005rt}.
Requiring the time dependence of the metric 
for general black $p$-brane systems, 
it was found that the 
structure of a warp factor that depends on the time is different from
the usual "product type" ansatz \cite{Duff:1986hr, Ohta:2006sw}.

In this paper, we consider dynamical solutions for intersecting 
brane systems in supergravity, in which 
gravity is not only coupled to a single gauge field but 
also to several combinations of scalar and gauge fields.
Although the configurations we consider are not directly related to
actual warped compactifications to four dimensions, we do
so in the hope that they may serve as a first step toward the 
understanding of realistic, dynamical compactifications that 
should have occurred in the early Universe.

Some intersecting brane solutions were originally founded by 
G\"{u}ven in eleven-dimensional supergravity~\cite{Gueven:1992hh}. 
A D4-D8 brane system was constructed by Polchinski and Witten
using the dual description 
of the type I string theory on ${\mathbb{R}}^9\times {\rm S}^1$ with 
$N$ coinciding D5 branes wrapping the circle \cite{Polchinski:1995df}.

After that, several authors investigated solutions describing
intersecting branes, and constructed new static 
solutions~\cite{Youm:1999zs,Ohta:1997gw}. 
Nastase analyzed a D4-D8 brane solution as 
a setup for the holographic dual of QCD~\cite{Nastase:2003dd}. 
Brandhuber and Oz found a classical solution of 
intersecting D4-D8 branes \cite{Brandhuber:1999np}.
In a certain region of spacetime,
the geometry is warped ${\rm AdS}_6 \times {\rm S}^4$,
which describes spontaneous compactification of 
the massive IIA supergravity in ten dimensions~\cite{Romans:1985tz}
to the gauged supergravity~\cite{Romans:1985tw, Cvetic:1999un}. 
In connection with gauge/gravity correspondence,
a supergravity background of D3-D7
brane intersection has been constructed which includes a flavor
D7 brane\cite{Aharony:1998xz, Grana:2001xn,
Burrington:2004id, Kirsch:2005uy}.

The paper is organized as follows. 
In \S Sec.\,\ref{sec:Dynamical p-brane solution},
we first consider $p$-brane systems in $D$ dimensions
and derive a class of dynamical solutions under a certain metric
ansatz.
In \S Sec.\,\ref{sec:Dynamical solution for D4-D8 brane system},
focusing on intersecting D4-D8 brane systems in the ten-dimensional 
type IIA supergravity, we extend the metric ansatz used in the
previous section to intersecting branes
and obtain a class of dynamical solutions.
Then further specializing the form of the metric, 
we consider a cosmological solution.
Interestingly, this solution is found to
approach a warped static solution as $\tau\to0$
and a Kasner-type anisotropic solution as $\tau\to\infty$,
where $\tau$ is the cosmic time.
We investigate the dynamical D3-D7 brane solution 
of the ten-dimensional type IIB supergravity
in \S Sec.\,\ref{sec:Dynamical D3-D7 brane solution}. 
Finally, we conclude in \S Sec.\,\ref{sec:Discussions}.


\section{Dynamical $p$-brane solutions}
\label{sec:Dynamical p-brane solution}

In this section, we consider dynamical $p$-brane systems in $D$
dimensions.
First, we write down the Einstein equations under a particular
ansatz for the metric, which is a generalization of
the metric form of known static $p$-brane solutions.
Then, we solve the Einstein equations
and present the solutions explicitly.

We consider a gravitational theory with the metric $g_{MN}$,
dilaton $\phi$, 
and an antisymmetric tensor field of rank $(p+2)$. 
This corresponds to a $p$-brane system in string theory.
The most general action for the $p$-brane system in the Einstein 
frame can be written as \cite{Lu:1995cs}
\Eq{
S=\frac{1}{2\kappa^2}\int \left(R\ast{\bf 1}_D
 -\frac{1}{2}d\phi \wedge \ast d\phi 
 -\frac{1}{2}e^{-c\phi}F_{(p+2)}\wedge\ast F_{(p+2)}\right),
\label{eq:sec2:D-dim action}
}
where $\kappa^2$ is the $D$-dimensional gravitational constant, 
$\ast$ is the Hodge dual operator in the $D$-dimensional spacetime,
and $c$ is a constant given by 
\Eq 
{
c^2=4-\frac{2(p+1)(D-p-3)}{D-2}.
   \label{eq:sec2:parameter c} 
   } 
The expectation values of fermionic fields are assumed to be zero.

After variations with respect to the metric, the dilaton, 
and the $(p+1)$-form gauge field, we obtain the field equations
\Eqr{
&&R_{MN}=\frac{1}{2}\pd_M\phi \pd_N \phi\nn\\
&&~+\frac{1}{2}\frac{1}{(p+2)!}e^{-c\phi} 
\left[(p+2)F_{MA_2\cdots A_{p+2}} {F_N}^{A_2\cdots A_{p+2}}
-\frac{p+1}{D-2}g_{MN} F^2_{(p+2)}\right],
   \label{eq:sec2:p-brane Einstein equation}\\
&&\triangle\phi=-\frac{1}{2}\frac{c}{(p+2)!}
e^{-c\phi}F^2_{(p+2)},
   \label{eq:sec2:p-brane scalar field equation}\\
&&d\left(e^{-c\phi}\ast F_{(p+2)}\right)=0,
   \label{eq:sec2:p-brane gauge field equation}
}
where $\triangle$ is the $D$-dimensional Laplace operator.

To solve the field equations, we assume the $D$-dimensional metric
in the form
\Eq{
ds^2=h^a(x, y)q_{\mu\nu}dx^{\mu}dx^{\nu} 
  +h^b(x, y)u_{ij}dy^idy^j,  
 \label{eq:sec2:metric of p-brane solution}
}
where $q_{\mu\nu}$ is a $(p+1)$-dimensional metric, which
depends only on the $(p+1)$-dimensional coordinates $x^{\mu}$, 
and $u_{ij}$ is the $(D-p-1)$-dimensional metric, which
depends only on the $(D-p-1)$-dimensional coordinates $y^i$. 
The parameters $a$ and $b$ are given by 
\Eq{
a=-\frac{D-p-3}{D-2},~~~~b=\frac{p+1}{D-2}.
 \label{eq:sec2:paremeter of the warp factor}
}
The metric form~(\ref{eq:sec2:metric of p-brane solution}) is
a straightforward generalization of the case of a static $p$-brane
system with a dilaton coupling \cite{Lu:1995cs}.
Furthermore, we assume that the scalar field $\phi$ and 
the gauge field strength $F_{(p+2)}$ are given by
\Eqr{
&&e^{\phi}=h^{-c/2},
  \label{eq:sec2:ansatz for the scalar field}\\
&&F_{(p+2)}=d(h^{-1})\wedge\Omega(\Xsp),
  \label{eq:sec2:ansatz for the gauge field}
}
where $\Omega(\Xsp)$ denotes the volume $(p+1)$ form
\Eq{
\Omega(\Xsp)=\sqrt{-q}\,dx^0\wedge dx^1\wedge \cdots \wedge 
dx^p.
}
Here, $q$ is the determinant of the metric $q_{\mu\nu}$.

Let us first consider the Einstein Eqs.
(\ref{eq:sec2:p-brane Einstein equation}). 
Using the assumptions (\ref{eq:sec2:metric of p-brane solution}), 
(\ref{eq:sec2:ansatz for the scalar field}) and 
(\ref{eq:sec2:ansatz for the gauge field}), 
the Einstein equations are given by
\Eqr{
&&R_{\mu\nu}(\Xsp)-h^{-1}D_{\mu}D_{\nu} h -\frac{a}{2}h^{-1} 
q_{\mu\nu}\left(\triangle_{\Xsp}h + h^{-1}\triangle_{\Ysp} h\right)=0,
 \label{eq:sec2:p-brane Einstein equation-mu}\\
&&R_{ij}(\Ysp)-\frac{b}{2} u_{ij}\left(\triangle_{\Xsp} h
 +h^{-1}\triangle_{\Ysp}h \right)=0,
 \label{eq:sec2:p-brane Einstein equation-ij}\\
&&\pd_{\mu}\pd_i h=0,
 \label{eq:sec2:p-brane Einstein equation-mi}
}
where $D_{\mu}$ is the covariant derivative with respect to 
the metric $q_{\mu\nu}$, 
$\triangle_{\Xsp}$ and $\triangle_{\Ysp}$ are 
the Laplace operators on the space of 
${\rm \Xsp}$ and the space ${\rm \Ysp}$, and 
$R_{\mu\nu}(\Xsp)$ and $R_{ij}(\Ysp)$ are the Ricci tensors
of the metrics $q_{\mu\nu}$ and $u_{ij}$, respectively.
From Eq.~(\ref{eq:sec2:p-brane Einstein equation-mi}), 
the warp factor $h$ must be in the form
\Eq{
h(x, y)= h_0(x)+h_1(y).
  \label{eq:sec2:form of warp factor}
}
With this form of $h$, the other components of
the Einstein Eqs.
~(\ref{eq:sec2:p-brane Einstein equation-mu})
and (\ref{eq:sec2:p-brane Einstein equation-ij}) 
are rewritten as
\Eqr{
&&R_{\mu\nu}(\Xsp)-h^{-1}D_{\mu}D_{\nu}h_0 -\frac{a}{2}h^{-1} 
q_{\mu\nu} \left( \triangle_{\Xsp} h_0  
+h^{-1}\triangle_{\Ysp}h_1\right)=0,
   \label{eq:sec2:p-brane Einstein equation-mu2}\\ 
&&R_{ij}(\Ysp)-\frac{b}{2}u_{ij}\left(\triangle_{\Xsp}h_0  
+h^{-1}\triangle_{\Ysp}h_1\right)=0. 
   \label{eq:sec2:p-brane Einstein equation-ij2}
   }

Let us next consider the gauge field.
Under the assumption (\ref{eq:sec2:ansatz for the gauge field}), 
we find
\Eq{ 
dF_{(p+2)}=h^{-1}(2\pd_i \ln h \pd_j \ln h + h^{-1}\pd_i\pd_j h)
dy^i\wedge dy^j\wedge\Omega(\Xsp) =0. 
}
Thus, the Bianchi identity is automatically satisfied.
Also the equation of motion for the gauge field becomes 
\Eqr{
d\left[e^{-c\phi}\ast F_{(p+2)} \right] 
=-d\left[\pd_ih (\ast_{\Ysp}dy^i)\right]=0,
 }
where $\ast_{\Ysp}$ denotes the Hodge dual operator on $\Ysp$.
Hence, the gauge field equation is automatically
satisfied under the assumption~(\ref{eq:sec2:ansatz for the gauge field}).

Let us consider the scalar field equation.
Substituting the forms of the scalar field 
(\ref{eq:sec2:ansatz for the scalar field}), 
the gauge field (\ref{eq:sec2:ansatz for the gauge field}), 
and the warp factor (\ref{eq:sec2:form of warp factor}) 
into the equation of motion for the scalar field  
(\ref{eq:sec2:p-brane scalar field equation}), we obtain
\Eq{
\frac{c}{2}h^{-b}\left(\triangle_{\Xsp}h_0
+h^{-1}\triangle_{\Ysp}h_1\right)=0,
  \label{eq:sec2:p-brane scalar field equation2}
}
Thus, unless the parameter $c$ is zero, the warp factor $h$ should
satisfy the equations
\Eq{
\triangle_{\Xsp}h_0=0, ~~~ \triangle_{\Ysp}h_1=0.
   \label{eq:sec2:solution for the scalar field}
}
If $F_{(p+2)}\ne 0$, the function $h_1$ is nontrivial.
In this case, 
the Einstein equations reduce to  
\Eqr{
&&R_{\mu\nu}(\Xsp)=0,
   \label{eq:sec2:Ricci tensor of p-brane solution}\\
&&R_{ij}(\Ysp)=0,\\
   \label{eq:sec2:Ricci tensor of p-brane solution2} 
&&D_{\mu}D_{\nu}h_0=0.
   \label{eq:sec2:warp factor of p-brane solution} 
 }
On the other hand, if $F_{(p+2)}=0$, the function $h_1$ becomes
trivial. Namely, the internal space is no longer
warped~\cite{Kodama:2005cz}.

As a special example, we consider the case
\Eq{ 
q_{\mu\nu}=\eta_{\mu\nu}\,,
\quad u_{ij}=\delta_{ij}\,,
 \label{eq:sec2:special metric of p-brane solution}
 }
where $\eta_{\mu\nu}$ is the $(p+1)$-dimensional 
Minkowski metric and $\delta_{ij}$ is 
the $(D-p-1)$-dimensional Euclidean metric. 
In this case, the solution for $h$ can be obtained
explicitly as
\Eq{ 
h(x, y)=A_{\mu}x^{\mu}+B
+\sum_{l}\frac{M_l}{|y^i-y^i_l|^{D-p-3}},
 \label{eq:sec2:special form of warp factor for p-brane solution}
}
where $A_{\mu}$, $B$, and $M_l$ are constant parameters. 
In the case of $D=11$, $p=2$, 
the solution describes an M2 brane 
\cite{Stelle:1998xg, Argurio:1998cp}.

Here, we mention an important fact about the nature of the
dynamical solutions described in the above. In general,
we regard the $(p+1)$-dimensional spacetime to contain
our four-dimensional universe, while the remaining space is
assumed to be compact and sufficiently small in size. Then 
one would usually think that an effective $(p+1)$-dimensional 
description of the theory should be possible at low energies. 
However, solutions of the above set of equations have the 
property that they are genuinely $D$ dimensional in the
sense that one can never neglect the dependence on $\Ysp$,
say of $h$. 
This is clear from an inspection of
Eqs.~(\ref{eq:sec2:p-brane Einstein equation-mu2}), 
(\ref{eq:sec2:p-brane Einstein equation-ij2}).
In particular, the second equation involves the Laplacian
of $h$ with respect to the space $\Xsp$. Hence, the equations
determining the internal space $\Ysp$ cannot be determined
independently from the geometry of the space $\Xsp$. 
The origin of this property is due to the existence of
a nontrivial gauge field strength, which forces the
function $h$ to be a linear combination of a function
of $x^\mu$ and a function of $y^i$, instead of a product
of these two types of functions as conventionally assumed.
This fact is in sharp contrast with the case when one
is allowed to integrate out the internal space to
obtain an effective lower-dimensional theory.

Finally, we comment on the exceptional case of $c=0$,
which happens when $(D,p)=(10,3)$, $(11,5)$, $(11,2)$.
The scalar field becomes constant because of
the ansatz (\ref{eq:sec2:ansatz for the scalar field}), and 
the scalar field Eq. 
(\ref{eq:sec2:p-brane scalar field equation2})
is automatically satisfied. 
Then, the Einstein equations become 
\Eqr{
&&R_{\mu\nu}(\Xsp)=0,
   \label{eq:sec2:Ricci tensor of p-brane solution c=0}\\
&&R_{ij}(\Ysp)=\frac{b}{2}(p+1)\lambda u_{ij}(\Ysp),\\
   \label{eq:sec2:Ricci tensor of p-brane solution2 c=0} 
&&D_{\mu}D_{\nu}h_0=\lambda q_{\mu\nu}(\Xsp),
   \label{eq:sec2:warp factor of p-brane solution c=0} 
 }
where $\lambda$ is a constant.
As seen from these equations, the internal space $\Ysp$ is not
necessarily Ricci flat, and the function $h_0$ becomes more complicated.
For example, when the metric $q_{\mu\nu}$ is Minkowski, $h_0$
is no longer linear in the coordinates $x^\mu$ but quadratic
in them \cite{Kodama:2005fz}.

Before concluding this section, 
let us study the special case of $(D,p)=(10,8)$ as a warm-up exercise
for the next section.
The existence of the D8 brane means the appearance of the 9-form 
gauge potential, hence the 10-form gauge field strength,
which is essentially a cosmological constant.
Then the ten-dimensional action in the Einstein frame
is given as \cite{Romans:1985tz}
\Eq{
S=\frac{1}{2\kappa^2}\int \left(R\ast{\bf 1}
 -\frac{1}{2}d\phi \wedge \ast d\phi 
 -\frac{1}{2}e^{5\phi/2}m^2 \ast{\bf 1}\right),
}
where $m$ is a constant parameter, which is
the dual of the 10-form field strength $F_{(10)}$ in the string frame.
This action is given by the truncations of Romans' massive 
type IIA supergravity 
\cite{Romans:1985tz, Bergshoeff:1996ui, Bergshoeff:1997ak, 
Imamura:2001cr}.

The field equations are expressed as
\Eqr{
\triangle\phi&=&\frac{5}{4}e^{5\phi/2}m^2\,,
\label{eq:sec2:massive-scalar}\\
R_{MN}&=&\frac{1}{2}\pd_M\phi \pd_N\phi+\frac{m^2}{16} 
   e^{5\phi/2}g_{MN}\,,
\label{eq:sec2:massive-Einstein}
}
where $\triangle$ is the ten-dimensional Laplace operator.
The ansatz for the metric~(\ref{eq:sec2:metric of p-brane solution}) 
and the scalar field~(\ref{eq:sec2:ansatz for the scalar field})
become
\Eqr{
ds^2&=& h^{1/8}(x, y)q_{\mu\nu}({\rm X})dx^{\mu}dx^{\nu}
             +h^{9/8}(x, y)dy^2,
             \label{eq:sec2:massive metric}\\
e^{\phi}&=&h^{-5/4}\,.
\label{eq:sec2:massive dilaton}
}
The field equations give
\Eqr{
&&R_{\mu\nu}({\rm X})=0,\\
&&h(x, y)=h_0(x)\pm m(y-y_0)
}
where $y_0$ is a constant parameter, and $h_0(x)$ has to satisfy the 
equation
\Eq{
D_{\mu}D_{\nu}h_0=0.
}


Now we show a solution for $h_0$ and 
$q_{\mu\nu}(\Xsp)$ to (\ref{eq:sec2:warp factor of p-brane solution}), 
(\ref{eq:sec2:warp factor of p-brane solution c=0}) 
except for $\Xsp$ being $(p+1)$-dimensional Minkowski spacetime. 
Let us consider a simple gravitational plane-wave metric
\Eq{
q_{\mu\nu}(\Xsp)dx^{\mu}dx^{\nu}
=dudv-K(u,x)du^2+\delta_{ij}dx^idx^j,
}
where $(i, j)=(2, \cdots, p+1)$.
This is a vacuum solution of the Einstein equations if the function 
$K$ satisfies
\Eq{
R_{uu}(\Xsp)=\frac{1}{2}\pd_i\pd^iK=0.
  \label{pw:Einstein equation:Eq}
}
Particular solutions are $K_1=a_{ij}(u)x^ix^j$ with $a_{ii}=0$, or 
$K_2=f(u)|x|^{-p+3}$ for $x\ne 0$.

First, we consider the $c\ne 0$ case in which $D_v h_0=D_ih_0=0$.
In this case, Eq.~(\ref{eq:sec2:warp factor of p-brane solution}) is 
\Eq{
\pd_u^2h_0=0.
}
Then, the form of $h_0$ is given by 
\Eq{
h_0(u)=c_1 u+c_2,
}
where $c_1$ and $c_2$ are constant parameters.
The metric of $D$-dimensional spacetime can be written as
\Eq{
ds^2=\left(c_1u+c_2+h_1\right)^aq_{\mu\nu}(\Xsp)dx^{\mu}dx^{\nu}
  +\left(c_1u+c_2+h_1\right)^bu_{ij}(\Ysp)dy^idy^j,
    }
where $h_1$ satisfies the equation $\lap_{\Ysp}h_1=0$.
Next, for the case $c=0$ in which $D_v h_0=D_ih_0=0$, 
Eq. 
(\ref{eq:sec2:warp factor of p-brane solution c=0}) becomes 
\Eq{
\pd_u^2h_0=\lambda. 
}
The solution of $h_0$ is
\Eq{
h_0(u)=\frac{\lambda}{2} u^2+c_1 u+c_2,
  \label{pp:h0:Eq}
}
where $c_1$ and $c_2$ are constant parameters.
The solution (\ref{pp:h0:Eq}) leads to the 
metric of $D$-dimensional spacetime   
\Eq{
ds^2=\left(\frac{\lambda}{2} u^2+c_1 u+c_2+h_1\right)^a
    q_{\mu\nu}(\Xsp)dx^{\mu}dx^{\nu}
    +\left(\frac{\lambda}{2} u^2
    +c_1 u+c_2+h_1\right)^b
    u_{ij}(\Ysp)dy^idy^j.
    }

Here, we mention that in the limit when the terms with $h_1$ dominates
in the metric the whole $D$-dimensional metric becomes static.
This is the metric of the static $p$-brane solution \cite{Lu:1995cs}, 
which is the so-called warped compactification. On the other hand, 
in the limit when the terms with $h_1$ are negligible,
the background changes from the above description to ordinary 
Kaluza-Klein compactification as time evolves.
Although this solution cannot describe 
a realistic cosmology, it is interesting to note that this cosmological 
solution is an asymptotically static $p$-brane solution.


\section{Dynamical solutions for the D4-D8 brane system}
\label{sec:Dynamical solution for D4-D8 brane system}

Now we consider dynamical solutions for the D4-D8 brane system, 
which appears in the ten-dimensional type IIA supergravity. 
The bosonic action of D4-D8 brane system in the 
Einstein frame is given by 
\cite{Youm:1999zs, Nastase:2003dd, Brandhuber:1999np, Cvetic:1999xx}
\Eq{
\hspace{-1cm}
S=\frac{1}{2{\kappa}^2}\int \left(R\ast{\bf 1}
 -\frac{1}{2}d\phi \wedge \ast d\phi 
 -\frac{1}{2\cdot 4!}e^{\phi/2}F_{(4)}\wedge\ast F_{(4)}
 -\frac{1}{2}e^{5\phi/2}m^2 \ast{\bf 1}\right).
   \label{eq:sec3:action of D4/D8}
   }
The equations of motion for $\phi$ and $F_{(4)}$,
and the Bianchi identity for $F_{(4)}$ are
\Eqr{
&&\triangle\phi=\frac{1}{4}\left(5m^2e^{5\phi/2}+
 \frac{1}{4!}e^{\phi/2}F_{(4)}^2\right),
  \label{eq:sec3:D4/D8 scalar field equation}\\
&&d(e^{\phi/2}\ast F_{(4)})=0\,,
  \label{eq:sec3:D4/D8 gauge field equation}\\
&&dF_{(4)}=0\,,
  \label{eq:sec3:D4/D8 Bianchi identity}
  }
where $\triangle$ is the ten-dimensional Laplace operator.
The Einstein equations are
\Eq{
\hspace{-2cm}
R_{MN}=\frac{1}{2}\pd_M\phi \pd_N\phi 
+ \frac{1}{16}m^2e^{5\phi/2}g_{MN}
+\frac{1}{2\cdot 4!}e^{\phi/2}\left(4F_{MABC}{F_N}^{ABC}
-\frac{3}{8}g_{MN}F_{(4)}^2\right).
   \label{eq:sec3:D4/D8 Einstein equations}
}

In the following, we look for a solution whose spacetime metric 
has the form
\Eqr{
&&\hspace{-1.5cm}ds^2=h^{1/12}(z)\left[h_4^{-3/8}(x,r,z)\,ds^2(\X5sp)
       +h_4^{5/8}(x,r,z)\left(dr^2+r^2 ds^2(\Y3sp)+dz^2\right)\right],
   \label{eq:sec3:D4/D8 ansatz for metric}\\
&&\hspace{-1.5cm}ds^2(\X5sp)=q_{\mu\nu}dx^{\mu}dx^{\nu}\,,\\
&&\hspace{-1.5cm}ds^2(\Y3sp)=u_{ij}dy^idy^j,
   \label{sec3:D4D8 metric:Eq}
}
where $q_{\mu\nu}$ is the five-dimensional metric 
depending only on the coordinates $x^\mu$ of $\X5sp$, and 
$u_{ij}$ is the three-dimensional metric depending only 
on the coordinates $y^i$ of $\Y3sp$. 
As for the scalar field and the 4-form field strength,
we adopt the following assumptions:
\Eqr{
e^{\phi}&=&h^{-5/6}h_4^{-1/4},
  \label{eq:sec3:D4/D8 ansatz for scalar field}\\
F_{(4)}&=&e^{-\phi/2}\ast \left[d(h_4^{-1})\wedge \Omega(\X5sp)\right],
  \label{eq:sec3:D4/D8 ansatz for gauge field}
}
where $\Omega(\X5sp)$ is given by
\Eq{
\Omega(\X5sp)=\sqrt{-q}\,dx^0\wedge dx^1\wedge dx^2 \wedge dx^3\wedge dx^4.
}

Let us first consider the Einstein 
Eqs. 
(\ref{eq:sec3:D4/D8 Einstein equations}). 
Under the assumptions (\ref{eq:sec3:D4/D8 ansatz for metric}), 
(\ref{eq:sec3:D4/D8 ansatz for scalar field}) 
and (\ref{eq:sec3:D4/D8 ansatz for gauge field}), 
the Einstein equations become 
\Eqr{
&&\hspace{-2cm}
R_{\mu\nu}(\X5sp) -h_4^{-1} D_{\mu}D_{\nu}h_4 
+ q_{\mu\nu}h_4^{-1}\left[\frac{3}{16}\triangle_{\X5sp}h_4
+\frac{3}{16}h_4^{-1}\pd_r^2 h_4-\frac{1}{24}h^{-1}\pd^2_z h
+\frac{3}{16}h_4^{-1}\pd_z^2 h_4
\right.\nn\\
 &&\left.~~~ 
+\frac{1}{36}\left(\pd_z \ln h\right)^2 
-\frac{m^2}{16} h^{-2} +\frac{1}{16}\pd_z\ln h\,\pd_z\ln h_4
+\frac{9}{16r} \pd_r \ln h_4 \right]=0,
  \label{eq:sec3:D4/D8 Einstein equation-mu}\\
&&\pd_{\mu}\pd_r h_4=0,
  \label{eq:sec3:D4/D8 Einstein equation-mr}\\
&&\pd_{\mu}\pd_z h_4=0,
  \label{eq:sec3:D4/D8 Einstein equation-mz}\\ 
&&\triangle_{\X5sp} h_4 + h_4^{-1}\left[ 
\pd_r^2 h_4 + \frac{3}{r}\pd_r h_4 +\pd_z^2 h_4 + \frac{1}{3}
\pd_z\ln h\, \pd_z h_4\right]+\frac{2}{15} h^{-1} \pd_z^2 h\nn\\
&& ~~~ +\frac{1}{5}\left[\frac{4}{9}\left(\pd_z\ln h\right)^2- 
m^2 h^{-2}\right]=0,
  \label{eq:sec3:D4/D8 Einstein equation-rr}\\
&&R_{ij}(\Y3sp)-2 u_{ij} -\frac{5}{16} r^2 u_{ij} 
\left[\triangle_{\X5sp}h_4 + h^{-1} \pd_r^2 h_4+\frac{3}{r}  
\pd_r \ln h_4 + h_4^{-1} \pd_z^2 h_4  \right.\nn\\
&& \left.~~~ + 
\frac{1}{3}\pd_z\ln h\, \pd_z\ln h_4 + \frac{1}{24}h^{-1}
\pd_z^2 h +\frac{1}{36}\left(\pd_z \ln h \right)^2-\frac{m^2}{16}h^{-2}
\right]=0,
  \label{eq:sec3:D4/D8 Einstein equation-ij}
} 
where $R_{\mu\nu}(\X5sp)$, $R_{ij}(\Y3sp)$ are the Ricci tensors 
of the metric $q_{\mu\nu}$ and $u_{ij}$, respectively, 
$D_{\mu}$ is the covariant derivative with respective to 
the metric $q_{\mu\nu}$, and 
$\triangle_{\X5sp}$ is the Laplace operator on the space 
$\X5sp$. 
{}From Eqs.~(\ref{eq:sec3:D4/D8 Einstein equation-mr})
and (\ref{eq:sec3:D4/D8 Einstein equation-mz}), 
the warp factor $h_4$ can be written as
\Eq{ 
h_4(x,y,z)=H_0(x) + H_1(r, z). 
   \label{eq:sec3:form of the h4-2}
}
Inserting this into Eqs.~(\ref{eq:sec3:D4/D8 Einstein equation-mu}), 
(\ref{eq:sec3:D4/D8 Einstein equation-rr}), and 
(\ref{eq:sec3:D4/D8 Einstein equation-ij}), we find
\Eqr{
&&\hspace{-2cm}R_{\mu\nu}(\X5sp) -h_4^{-1} D_{\mu}D_{\nu}H_0 
+\frac{3}{16}h_4^{-1}q_{\mu\nu}\left[
\frac{1}{3}\left(
\frac{4}{9}\left(\pd_z \ln h\right)^2 
-\frac{2}{9}h^{-1}\pd^2_z h
-m^2h^{-2} \right)
\right.\nn\\
&&\hspace{-2cm}~~
\left.
+\triangle_{\X5sp}H_0
+h_4^{-1}\left(\pd_r^2 H_1+\frac{3}{r}\pd_r H_1 
+\pd_z^2 H_1+\frac{1}{3}\pd_z\ln h\,\pd_zH_1\right)
\right]=0\,,
  \label{eq:sec3:D4/D8 Einstein equation-mu2}\\
&&\hspace{-2cm}
\triangle_{\X5sp} H_0 + h_4^{-1}\left( 
\pd_r^2 H_1 + \frac{3}{r}\pd_r H_1 +\pd_z^2 H_1 + \frac{1}{3}
\pd_z\ln h\, \pd_z H_1\right)
\nn\\
&&\hspace{-2cm}~~
+\frac{1}{5}\left(
\frac{4}{9}\left(\pd_z\ln h\right)^2
+\frac{2}{3}h^{-1} \pd_z^2 h
-m^2 h^{-2}\right)=0\,,
  \label{eq:sec3:D4/D8 Einstein equation-rr2}\\
&&\hspace{-2cm}
R_{ij}(\Y3sp)-2 u_{ij} -\frac{5}{16} r^2 u_{ij} 
\left[\triangle_{\X5sp}H_0 + 
h_4^{-1}
\left(\pd_r^2 H_1+\frac{3}{r}\pd_r H_1 +\pd_z^2 H_1
+\frac{1}{3}\pd_z\ln h\, \pd_zH_1 \right)
 \right.\nn\\
&&\hspace{-2cm}~~
\left.
+ \frac{1}{16}
\left(\frac{4}{9}\left(\pd_z \ln h \right)^2
+\frac{2}{3}h^{-1}\pd_z^2 h
-m^2h^{-2}\right)
\right]=0\,.
  \label{eq:sec3:D4/D8 Einstein equation-ij2}
} 

Let us next consider the gauge field equations.
The gauge field 
Eq.~(\ref{eq:sec3:D4/D8 gauge field equation}) 
is automatically satisfied under the 
assumption~(\ref{eq:sec3:D4/D8 ansatz for gauge field}) 
and the form of $h_4$ given by Eq.~(\ref{eq:sec3:form of the h4-2}).
Under the assumptions (\ref{eq:sec3:D4/D8 ansatz for metric}) and 
(\ref{eq:sec3:D4/D8 ansatz for scalar field}), 
(\ref{eq:sec3:D4/D8 ansatz for gauge field}), the Bianchi identity 
(\ref{eq:sec3:D4/D8 Bianchi identity}) gives
\Eqr{
&&\pd_r^2 h_4+\frac{3}{r}\pd_rh_4+\pd_z^2h_4
+\frac{1}{3}\pd_z\ln h\,\pd_zh_4=0\,,
  \label{eq:sec3:D4/D8 Bianchi identity1}\\
&&\pd_{\mu}\pd_rh_4=0\,,
  \label{eq:sec3:D4/D8 Bianchi identity2}\\
&&\pd_{\mu}\pd_zh_4=0\,.
  \label{eq:sec3:D4/D8 Bianchi identity3}
}
The last two equations are the same as 
Eqs.~(\ref{eq:sec3:D4/D8 Einstein equation-mr}) 
and (\ref{eq:sec3:D4/D8 Einstein equation-mz}),
and they have been already solved to give
Eq.~(\ref{eq:sec3:form of the h4-2}). 
Then the first 
Eq.~(\ref{eq:sec3:D4/D8 Bianchi identity1}) becomes
\Eq{
\pd_r^2 H_1+\frac{3}{r}\pd_rH_1+\pd_z^2H_1
+\frac{1}{3}\pd_z\ln h\,\pd_zH_1=0\,.
  \label{eq:sec3:D4/D8 Bianchi identity4}
}

Next we consider the scalar field 
Eq.~(\ref{eq:sec3:D4/D8 scalar field equation}). 
Substituting the assumptions for the metric 
(\ref{eq:sec3:D4/D8 ansatz for metric}), the
scalar and gauge fields~
(\ref{eq:sec3:D4/D8 ansatz for scalar field}), 
(\ref{eq:sec3:D4/D8 ansatz for gauge field}), 
and the form of $h_4$ (\ref{eq:sec3:form of the h4-2}) into 
the scalar field 
Eq.~(\ref{eq:sec3:D4/D8 scalar field equation}), 
we find
\Eqr{
&&h^{-1/12}h_4^{-5/8}\left[\triangle_{\X5sp}H_0
+\frac{5}{9}\left(\pd_z\ln h\right)^2-\frac{5}{4}m^2h^{-2}
+\frac{5}{6}h^{-1}\pd_z^2h\right.
\nn\\
&&\qquad\qquad
\left.
+h_4^{-1}\left(\pd_r^2 H_1+\frac{3}{r}\pd_rH_1
+\pd_z^2H_1+\frac{1}{3}\pd_z\ln h\,\pd_z H_1\right)\right]=0\,.
   \label{eq:sec3:D4/D8 scalar field equation2}
}
Together with Eq.~(\ref{eq:sec3:D4/D8 Bianchi identity4}),
the above equation gives
\Eqr{
&&\triangle_{\X5sp}H_0
+\frac{5}{4}
\left(\frac{4}{9}\left(\pd_z\ln h\right)^2
+\frac{2}{3}h^{-1}\pd_z^2h
-m^2h^{-2}
\right)=0\,.
   \label{eq:sec3:reduced field equations1}
   }

Inserting Eqs.~(\ref{eq:sec3:D4/D8 Bianchi identity4})
 and (\ref{eq:sec3:reduced field equations1}) into the 
Einstein 
Eqs.~(\ref{eq:sec3:D4/D8 Einstein equation-mu2}) - 
(\ref{eq:sec3:D4/D8 Einstein equation-ij2}),
we find for nontrivial $H_1$,
\Eqr{
&&R_{\mu\nu}(\X5sp)=0,
  \label{eq:sec3:Einstein equations for D4D8 1}\\
&&R_{ij}(\Y3sp)=2 u_{ij},
  \label{eq:sec3:Einstein equations for D4D8 2}\\
&&D_{\mu}D_{\nu}H_0=0\,,
\quad\triangle_{\X5sp}H_0=0\,,
  \label{eq:sec3:Einstein equations for D4D8 3}\\
&&\frac{4}{9}\left(\pd_z h\right)^2-m^2=0\,,
\quad \pd_z^2h=0\,.
 \label{eq:sec3:reduced field equations4}
}
The last line of the above equations is immediately solved to give
\Eq{
h(z)=\frac{3}{2}m(z-z_0)\,,
  \label{eq:sec3:solution of field equations h}
}
where $z_0$ is an integration constant
(corresponding to the position of the D8 brane).
Below we set $z_0=0$ without loss of generality.
Then Eq.~(\ref{eq:sec3:D4/D8 Bianchi identity4}) reduces to
\begin{eqnarray}
\pd_r^2 H_1+\frac{3}{r}\pd_rH_1+\pd_z^2H_1
+\frac{1}{3z}\,\pd_zH_1=0\,.
\end{eqnarray}
The solution is
\Eq{
H_1(r, z)=\frac{c_1}{(r^2+z^2)^{5/3}}+c_2\,,
\label{eq:sec3:solution of field equations H1}
}
where $c_1$ and $c_2$ are constant parameters.

Let us investigate the geometrical properties of 
the D4-D8 brane system. 
As a particular solution to the three-dimensional
metric $u_{ij}$ which satisfies
Eq.~(\ref{eq:sec3:Einstein equations for D4D8 2}),
we take the space $\Y3sp$ to be a three-dimensional sphere ${\rm S}^3$.
Then if we make a change of coordinates, 
$z=\tilde{r}\sin\alpha$, $r=\tilde{r}\cos\alpha$ ($0\le\alpha\le\pi/2$),
the metric reads
\Eq{
ds^2=h^{1/12}\left[h_4^{-3/8}q_{\mu\nu}dx^{\mu}dx^{\nu}
+h_4^{5/8}(d\tilde{r}^2+\tilde{r}^2d\Omega_4^2)\right],
   \label{eq:sec3:metric of near horizon limit1}
}
where 
\Eqr{
&&d\Omega_4^2=d\alpha^2+\cos^2\alpha d\Omega_3^2\,,\\
&&h_4(x, \tilde{r})=H_0(x)+\frac{c_1}{\tilde{r}^{10/3}}+c_2
=H_0(x)+H_1(\tilde{r})\,,\\
&&h(\tilde{r}, \alpha)=\frac{3}{2}m\tilde{r}\sin\alpha\,.
}
Here, $d\Omega_3^2$ and $d\Omega_4^2$ denote the line elements 
of the three-dimensional sphere ${\rm S}^3$
and the four-dimensional sphere ${\rm S}^4$, respectively. 

Now we further define a new coordinate $U$ by $\tilde{r}^2=U^3$.
 From Eq.~(\ref{eq:sec3:Einstein equations for D4D8 3}),
we see that $H_0$ is a linear function of $x^\mu$. Hence,
keeping the values of these coordinates finite, 
the metric in the limit $U\rightarrow 0$ becomes 
\Eq{
ds^2=\left(\frac{3}{2}m\sin\alpha\right)^{1/12}
\left[c_1^{-1/2}U^2q_{\mu\nu}dx^{\mu}dx^{\nu}
+c_1^{1/2}\left(\frac{9dU^2}{4U^2}+d\Omega_4^2\right)\right],
   \label{eq:sec3:metric of near horizon limit2}
} 
while the dilaton is given by 
\Eq{
e^{\phi}=c_1^{-1/4}\left(\frac{3}{2}m\sin\alpha\right)^{-5/6}.
}
This is a static metric.
In particular, in the case $q_{\mu\nu}$ is the five-dimensional 
Minkowski metric $\eta_{\mu\nu}$, the above ten-dimensional 
metric becomes a warped ${\rm AdS}_6\times {\rm S}^4$ space 
\cite{Brandhuber:1999np, Cvetic:1999un, Behrndt:1999mk, Nunez:2001pt}.

Let us consider the case $q_{\mu\nu}=\eta_{\mu\nu}$ in more detail.
In this case, a solution for the warp factors $h_4$ and $h$ can be obtained
explicitly as
\Eqr{ 
h_4(x, \tilde{r})&=\beta t + K,\\
K&\equiv&\beta_a x^a+\gamma+H_1(\tilde{r}),\\
h(\tilde{r}, \alpha)&=&\frac{3}{2}m\tilde{r}\sin\alpha\,,
 \label{eq:sec2:special form of warp factor for D4/D8-brane solution}
}
where $x^a~(a=1, \cdots, 4)$ denote the spatial coordinates of
${\rm X}_5$, $H_1(\tilde{r})$ is given by
(\ref{eq:sec3:solution of field equations H1}), and 
$\beta$, $\beta_{a}$ and $\gamma$ are constant parameters.

Now we introduce a new time coordinate $\tau$ by
\Eq{
\frac{\tau}{\tau_0}=(\beta t)^{13/16}\,,
\qquad \beta\tau_0=\frac{16}{13}\,,
}
where we have assumed $\beta>0$ for simplicity.
Then, the ten-dimensional metric 
(\ref{eq:sec3:metric of near horizon limit1}) 
is given by
\Eqr{ 
ds^2&=&h^{1/12}
\left(1+\left(\frac{\tau}{\tau_0}\right)^{-16/13}K\right)^{-3/8}
\left[
\left(-d\tau^2+\left(\frac{\tau}{\tau_0}\right)^{-6/13}
\delta_{ab}dx^adx^b\right)
\right.\nn\\
&&\left.
+\left(1+\left(\frac{\tau}{\tau_0}\right)^{-16/13}K\right)^{5/8}
\left(\frac{\tau}{\tau_0}\right)^{10/13}
\left(d\tilde{r}^2+\tilde{r}^2d\Omega_4^2\right)\right]\,,
 \label{eq:sec3:special metric of D4/D8-brane solution conformal time}
 }
where the metric $\delta_{ab}$ is the spatial part of the five-dimensional
Minkowski metric $\eta_{\mu\nu}$.
If we set $K=0$, the scale factor of the four-dimensional space
is proportional to $\tau^{-6/13}$, while that for the remaining
five-dimensional space is proportional to $\tau^{10/13}$.
Thus, in the limit when the terms with $K$ are negligible, 
which is realized in the limit $\tau\to\infty$, we have
a Kaluza-Klein type cosmological solution. 
Again, similar to the case of the $p$-brane solution discussed 
in Sec.~\ref{sec:Dynamical p-brane solution},
although this solution is by no means realistic, 
it is interesting to note that it is asymptotically static in the past
$\tau\to0$.


\section{Dynamical D3-D7 brane solution}
\label{sec:Dynamical D3-D7 brane solution}

In this section, we consider dynamical solutions for the D3-D7 
brane system, which appears in the ten-dimensional type IIB 
supergravity \cite{Schwarz:1983qr}.

\subsection{D7-brane solution}
\label{subsec:D7-brane}
First, we discuss a D7-brane solution \cite{Greene:1989ya}.
The action for the D7-brane system in the Einstein 
frame can be written as 
\Eq{
S=\frac{1}{2\kappa^2}\int \left(R\ast{\bf 1}
 -\frac{1}{2({\rm Im}\,\tau)^2}d\tau \wedge \ast d\bar{\tau} \right),
\label{D7CS:10-dim action:Eq}
}
where $\kappa^2$ is the ten-dimensional gravitational constant, 
$\ast$ is the Hodge dual operator in the ten-dimensional spacetime, and 
$\tau=C_{(0)}+ie^{-\phi}$.
The expectation values of fermionic fields are assumed to be zero.
We can write the field equations by using the ten-dimensional action 
(\ref{D7CS:10-dim action:Eq}),
\Eqr{
&&R_{MN}=\frac{1}{({\rm Im}\,\tau)^2}
   \left(\pd_M\tau \pd_N \bar{\tau}+\pd_M\bar{\tau} \pd_N \tau\right),
   \label{D7CS:D7-brane Einstein equation:Eq}\\
&&\triangle\tau=\frac{1}{i({\rm Im}\,\tau)}g^{MN}\pd_M\tau\pd_N\tau,
   \label{D7CS:D7-brane scalar field equation:Eq}
}
where $\triangle$ is the ten-dimensional Laplace operator.

We adopt the following ansatz for the ten-dimensional metric:
\Eq{
ds^2=q_{\mu\nu}(\Xsp)dx^{\mu}dx^{\nu} 
  +e^{\Psi(x, y)}\delta_{ij}(\Ysp)dy^idy^j,  
 \label{D7CS:metric of D7-brane solution:Eq}
}
where $q_{\mu\nu}$ is a eight-dimensional metric which
depends only on the eight-dimensional coordinates $x^{\mu}$, 
and $y^i$ denote the two-dimensional coordinates. 

We first consider the Einstein 
Eqs.~(\ref{D7CS:D7-brane Einstein equation:Eq}). 
Using the assumption (\ref{D7CS:metric of D7-brane solution:Eq}),
the Einstein equations are given by
\Eqr{
&&\hspace{-1cm}
R_{\mu\nu}(\Xsp)-D_{\mu}D_{\nu}\Psi 
  -\frac{1}{2}\pd_{\mu}\Psi\pd_{\nu}\Psi
  -\frac{1}{4({\rm Im}\tau)^2}\left(\pd_{\mu}\tau \pd_{\nu} \bar{\tau}
  +\pd_{\mu}\bar{\tau} \pd_{\nu} \tau\right)=0,
 \label{D7CS:D7-brane Einstein equation-mu:Eq}\\
&&\hspace{-1cm}
 \delta_{ij}e^{\Psi}\left(\triangle_{\Xsp} \Psi
 +q^{\rho\sigma}\pd_{\rho}\Psi\pd_{\sigma}\Psi\right)
 +\delta_{ij}\triangle_{\Ysp}\Psi+\frac{1}{2({\rm Im}\,\tau)^2}
 \left(\pd_i\tau \pd_j \bar{\tau}+\pd_i\bar{\tau} \pd_j \tau\right)=0,
 \label{D7CS:D7-brane Einstein equation-ij:Eq}\\
&&\hspace{-1cm}
\pd_{\mu}\pd_i \Psi+\frac{1}{2({\rm Im}\,\tau)^2}
 \left(\pd_{\mu}\tau \pd_i \bar{\tau}+\pd_{\mu}\bar{\tau} \pd_i \tau\right)=0,
 \label{D7CS:D7-brane Einstein equation-mi:Eq}
}
where $D_{\mu}$ is the covariant derivative with respect to 
the metric $q_{\mu\nu}(\Xsp)$, 
$\triangle_{\Xsp}$ and $\triangle_{\Ysp}$ are 
the Laplace operators on the space of 
${\rm \Xsp}$ and the space of ${\rm \Ysp}$, and 
$R_{\mu\nu}(\Xsp)$ is the Ricci tensors
of the metrics $q_{\mu\nu}$, respectively.

Now we introduce the complex coordinate $z=y^1+iy^2$ and 
assume that the scalar field $\tau$ depends only on the 
coordinate $z$ and $\bar{z}$ \cite{Greene:1989ya}
\Eq{
\tau=\tau(z, \bar{z}).
  \label{D7CS:ansatz for the scalar field:Eq}
}
Then, from \eqref{D7CS:D7-brane Einstein equation-mi:Eq}, 
the warp factor $\Psi$ must be in the form
\Eq{
\Psi(x, z, \bar{z})= \Psi_0(x)+\Psi_1(z, \bar{z}).
  \label{D7CS:form of warp factor:Eq}
}

Using the form of $\Psi$ and the assumption 
(\ref{D7CS:ansatz for the scalar field:Eq}), the field equations 
Eqs.~(\ref{D7CS:D7-brane Einstein equation-mu:Eq}) and 
(\ref{D7CS:D7-brane Einstein equation-ij:Eq}) 
are rewritten as
\Eqr{
&&R_{\mu\nu}(\Xsp)-D_{\mu}D_{\nu}\Psi_0 
  -\frac{1}{2}\pd_{\mu}\Psi_0\pd_{\nu}\Psi_0=0,
 \label{D7CS:D7-brane Einstein equation-mu2:Eq}\\
&&q^{\rho\sigma}(\Xsp)\left(D_{\rho}D_{\sigma}\Psi_0+
  \pd_{\rho}\Psi_0\pd_{\sigma}\Psi_0\right)=0,
 \label{D7CS:D7-brane Einstein equation-ij2:Eq}\\
&&\pd\bar{\pd}\Psi_1- \pd\bar{\pd}\ln({\rm Im}\,\tau) =0,
 \label{D7CS:D7-brane Einstein equation-ij2-2:Eq}
   }
where $\pd$, $\bar{\pd}$ are defined by
\Eq{
\pd=\frac{\pd}{\pd z}\equiv
\frac{1}{2}\left(\frac{\pd}{\pd y^1}-i\frac{\pd}{\pd y^2}\right),
~~~~\bar{\pd}=\frac{\pd}{\pd \bar{z}}\equiv
  \frac{1}{2}\left(\frac{\pd}{\pd y^1}+i\frac{\pd}{\pd y^2}\right).
}
If the function $\Psi_0$ satisfies the equation,
\Eq{
D_{\mu}D_{\nu}\Psi_0+\pd_{\mu}\Psi_0\pd_{\nu}\Psi_0=0,
}
Eq.~(\ref{D7CS:D7-brane Einstein equation-ij2:Eq}) is satisfied
and the field 
Eqs.~(\ref{D7CS:D7-brane Einstein equation-mu2:Eq}) and  
(\ref{D7CS:D7-brane Einstein equation-ij2-2:Eq})  
become
\Eqr{
&&R_{\mu\nu}(\Xsp)+\frac{1}{2}\pd_{\mu}\Psi_0\pd_{\nu}\Psi_0=
R_{\mu\nu}(\Xsp)-\frac{1}{2}D_{\mu}D_{\nu}\Psi_0=0,
 \label{D7CS:D7-brane Einstein equation-mu3:Eq}\\
&&\pd\bar{\pd}\Psi_1- \pd\bar{\pd}\ln({\rm Im}\,\tau) =0.
 \label{D7CS:D7-brane Einstein equation-ij3:Eq}
   }
If the eight-dimensional space X is assumed to be an
Einstein manifold, \eqref{D7CS:D7-brane Einstein equation-mu3:Eq}
implies that the eight-dimensional metric $q_{\mu\nu}$ 
is expressed as a product of two vectors. Hence, 
the determinant of the metric $q_{\mu\nu}$ becomes zero,
which is not permissible. It then follows that 
$R_{\mu\nu}(\Xsp)=0$ and $\Psi_0$ is constant,
which implies that the function $\Psi$ depends only on
the coordinates $z,~\bar{z}$. That is, the solution is static.
Static solutions of the field 
Eq.~(\ref{D7CS:D7-brane Einstein equation-ij3:Eq}) 
were discussed in \cite{Greene:1989ya, Aharony:1998xz, 
Burrington:2004id, Kirsch:2005uy}. 
The D7-brane solution (\ref{D7CS:metric of D7-brane solution:Eq})  
 is different from the $p$-brane solutions 
(\ref{eq:sec2:metric of p-brane solution}) 
because the scalar field $\tau$ does not depend on the 
eight-dimensional coordinate $x^{\mu}$. 
To obtain a time-dependent solution, we will 
add a D3 brane, which we will discuss in the next subsection. 

\subsection{D3-D7 brane solution}

We consider a gravitational theory with the metric $g_{MN}$,
scalar field $\tau$, and a 5-form field strength $F_{(5)}$. 
The action for the D3-D7 brane system in the Einstein 
frame can be written as
\Eq{
S=\frac{1}{2\kappa^2}\int \left(R\ast{\bf 1}
 -\frac{1}{2({\rm Im}\,\tau)^2}d\tau \wedge \ast d\bar{\tau} 
 -\frac{1}{4}F_{(5)}\wedge\ast F_{(5)}\right),
\label{D3D7-2:10-dim action:Eq}
}
where $\kappa^2$ is 
the ten-dimensional gravitational constant and
$\ast$ is the Hodge dual operator in the ten-dimensional spacetime.
The expectation values of fermionic fields are assumed to be zero.

The ten-dimensional action (\ref{D3D7-2:10-dim action:Eq}) 
gives following field equations:
\Eqr{
&&R_{MN}=\frac{1}{4({\rm Im}\,\tau)^2}
   \left(\pd_M\tau \pd_N \bar{\tau}+\pd_M\bar{\tau} \pd_N \tau\right)
+\frac{1}{96}F_{MABCD} {F_N}^{ABCD},
   \label{D3D7-2:Einstein equation:Eq}\\
&&\triangle\tau=\frac{1}{i({\rm Im}\,\tau)}g^{MN}\pd_M\tau\pd_N\tau,
   \label{D3D7-2:scalar field equation:Eq}\\
&&dF_{(5)}=0, ~~~F_{(5)}=\ast F_{(5)},
   \label{D3D7-2:gauge field equation:Eq}
}
where $\triangle$ is the ten-dimensional Laplace operator, and 
we used the self-duality condition for the 5-form field strength, 
which is required by supersymmetry \cite{Schwarz:1983qr}, and 
$\triangle$ is the ten-dimensional Laplace operator.

We assume the form of the ten-dimensional metric as
\Eqr{
ds^2&=&g_{MN}dx^Mdx^N\nn\\
    &=&h^{-1/2}(x, y)q_{\mu\nu}(\Xsp)dx^{\mu}dx^{\nu}
    +h^{1/2}(x, y)u_{ij}(\Ysp)dy^idy^j,
\label{D3D7-2:metric of solution:Eq}
}
where $q_{\mu\nu}(\Xsp)$ denotes a four-dimensional metric, which
depends only on the four-dimensional coordinates $x^{\mu}$, 
and $u_{ij}(\Ysp)$ is the six-dimensional metric, which
depends only on the six-dimensional coordinates $y^i$. 
The brane configuration is given as follows:
\begin{center}
\begin{tabular}{|c|c|c|c|c|c|c|c|c|c|c|}
\hline
&0&1&2&3&4&5&6&7&8&9\\
\hline
D3 & $\circ$ & $\circ$ & $\circ$ & $\circ$ &&&&&& \\
\hline
D7 & $\circ$ & $\circ$ & $\circ$ & $\circ$ & $\circ$ & $\circ$ 
& $\circ$ & $\circ$ & & \\ 
\hline
\end{tabular}
\end{center}

Furthermore, we assume that the scalar field $\tau$ and 
the gauge field strength $F_{5}$
are given by
\Eqr{
\tau&=&\tau(y),
  \label{D3D7-2:ansatz for the scalar field:Eq}\\
F_{(5)}&=&(1\pm\ast)d(h^{-1})\wedge\Omega(\Xsp),
  \label{D3D7-2:ansatz for the gauge field:Eq}
}
where $\Omega(\Xsp)$ denotes the volume 4-form
\Eq{
\Omega(\Xsp)=\sqrt{-q}dx^0\wedge dx^1\wedge dx^2\wedge 
dx^3.
}
Here, $q$ is the determinant of the metric $q_{\mu\nu}(\Xsp)$.

Now we assume that the 
metric $u_{ij}(\Ysp)$ is given by
\Eq{
u_{ij}(\Ysp)dy^idy^j=s_{ab}(\Ysp)dw^a dw^b+e^{\Psi(r)}(dr^2+r^2d\theta^2),
\label{D3D7-2:metric of internal space:Eq}
}
where $s_{ab}(\Ysp)$ is the metric of the four-dimensional space, 
and $\Psi$ is a function that depends only on the coordinate $r$.

We adopt an assumption for the scalar field $\tau$ and the function 
$\Psi$ as\cite{Greene:1989ya, Kirsch:2005uy}
\Eqr{
\tau&=&C_{(0)}+ie^{-\phi}\,;
\cr
C_{(0)}&=&A\theta,
   \label{D3D7-2:ansatz for scalar fields:Eq}\\
e^{-\phi(r)}&=&e^{\Psi(r)}=-A\ln\left(\frac{r}{r_{\rm c}}\right),
  \label{D3D7-2:ansatz for scalar fields2:Eq}
}
where $A$ and $r_{\rm c}$ are constant parameters.
We note that, in this case, the square of the Ricci tensor for 
the internal space (\ref{D3D7-2:metric of internal space:Eq}) is
given by 
\Eq{
R^{ij}(\Ysp)R_{ij}(\Ysp)=\frac{1}{2r^4}
\left[A\ln\left(\frac{r}{r_{\rm c}}\right)\right]^{-6}.
}
Thus, the six-dimensional space $\Ysp$ has singularities 
at $r=0$ and $r=r_{\rm c}$.

Let us first consider the Einstein 
Eqs.~(\ref{D3D7-2:Einstein equation:Eq}). 
Using the assumptions (\ref{D3D7-2:metric of solution:Eq}), 
(\ref{D3D7-2:ansatz for the scalar field:Eq}),  
(\ref{D3D7-2:ansatz for the gauge field:Eq}), 
(\ref{D3D7-2:metric of internal space:Eq}), 
the Einstein equations are given by
\Eqr{
&&R_{\mu\nu}(\Xsp)-h^{-1}D_{\mu}D_{\nu} h +\frac{1}{4}h^{-1} 
q_{\mu\nu}\left(\triangle_{\Xsp}h + h^{-1}\triangle_{\Ysp} h\right)=0,
 \label{D3D7-2:D3-D7 brane Einstein equation-mu:Eq}\\
&&R_{ij}(\Ysp)-\frac{1}{4} u_{ij}\left(\triangle_{\Xsp} h
 +h^{-1}\triangle_{\Ysp}h \right)=\frac{1}{4({\rm Im}\,\tau)^2}
   \left(\pd_i\tau \pd_j \bar{\tau}+\pd_i\bar{\tau} \pd_j \tau\right),
 \label{D3D7-2:D3-D7 brane Einstein equation-ij:Eq}\\
&&\pd_{\mu}\pd_i h=0,
 \label{D3D7-2:D3-D7 brane Einstein equation-mi:Eq}
}
where $D_{\mu}$ is the covariant derivative with respect to 
the metric $q_{\mu\nu}$, 
$\triangle_{\Xsp}$ and $\triangle_{\Ysp}$ are 
the Laplace operators on the space of 
${\rm \Xsp}$ and the space ${\rm \Ysp}$, respectively, and 
$R_{\mu\nu}(\Xsp)$ and $R_{ij}(\Ysp)$ are the Ricci tensors
of the metrics $q_{\mu\nu}$ and $u_{ij}$, respectively.
From \eqref{D3D7-2:D3-D7 brane Einstein equation-mi:Eq}, 
the warp factor $h$ must be in the form 
\Eq{
h(x, y)= h_0(x)+h_1(y).
  \label{D3D7-2:form of warp factor:Eq}
}
Using (\ref{D3D7-2:form of warp factor:Eq}),
Eqs.~(\ref{D3D7-2:D3-D7 brane Einstein equation-mu:Eq}) 
and (\ref{D3D7-2:D3-D7 brane Einstein equation-ij:Eq}) 
are rewritten as
\Eqr{
&&R_{\mu\nu}(\Xsp)-h^{-1}D_{\mu}D_{\nu}h_0 +\frac{1}{4}h^{-1} 
q_{\mu\nu} \left( \triangle_{\Xsp} h_0  
+h^{-1}\triangle_{\Ysp}h_1\right)=0,
   \label{D3D7-2:D3-D7 brane Einstein equation-mu2:Eq}\\ 
&&R_{ij}(\Ysp)-\frac{1}{4}u_{ij}\left(\triangle_{\Xsp}h_0  
+h^{-1}\triangle_{\Ysp}h_1\right)=\frac{1}{4({\rm Im}\,\tau)^2}
   \left(\pd_i\tau \pd_j \bar{\tau}+\pd_i\bar{\tau} \pd_j \tau\right). 
   \label{D3D7-2:D3-D7 brane Einstein equation-ij2:Eq}
   }
In order to simplify these equations,
we use the gauge and scalar field equations.

First we consider the gauge field equation.
Under the assumption (\ref{D3D7-2:ansatz for the gauge field:Eq}), 
we find
\Eqr{
dF_{(5)}=\mp \triangle_{\Ysp}h_1 \,dy^i\wedge \ast_{\Ysp} dy^i\nn
=0,
}
where we have used (\ref{D3D7-2:form of warp factor:Eq}), 
and $\ast_{\Ysp}$ denotes the Hodge dual operator on $\Ysp$.
Thus, we obtain the equation,
\Eq{
\lap_{\Ysp}h_1=0.
 \label{D3D7-2:solution of gauge fields equation:Eq}
}

Next we consider the scalar field equation.
Substituting the form of the scalar field,  
(\ref{D3D7-2:ansatz for scalar fields:Eq}) and 
(\ref{D3D7-2:ansatz for scalar fields2:Eq}), 
and the metric (\ref{D3D7-2:metric of solution:Eq}) with
(\ref{D3D7-2:metric of internal space:Eq}) 
into the equation of motion for the scalar field $\tau$, 
(\ref{D3D7-2:scalar field equation:Eq}),
one finds it is automatically satisfied.

Then, noting the fact that $\tau$ is a function of
only the coordinates $y^i$,
Eq.~(\ref{D3D7-2:D3-D7 brane Einstein equation-ij2:Eq})
implies $\lap_{\Xsp}h_0=0$ if the function $h_1$ is nontrivial.
In this case, the Einstein equations reduce to  
\Eqr{
&&R_{\mu\nu}(\Xsp)=0,
   \label{D3D7-2:Ricci tensor of D3-D7 brane solution:Eq}\\
&&R_{ab}(\Ysp)=0,
   \label{D3D7-2:Ricci tensor of D3-D7 brane solution2:Eq} \\
&&D_{\mu}D_{\nu}h_0=0,
   \label{D3D7-2:warp factor of D3-D7 brane solution:Eq} 
 }
where $R_{ab}(\Ysp)$ is the Ricci tensor with respect to the 
metric $s_{ab}(\Ysp)$.
The Einstein equations with respect to 
$R_{rr}$, $R_{r\theta}$, $R_{\theta\theta}$ are automatically 
satisfied under the 
assumptions (\ref{D3D7-2:metric of internal space:Eq}), 
(\ref{D3D7-2:ansatz for scalar fields:Eq}), 
(\ref{D3D7-2:ansatz for scalar fields2:Eq}). 
If the function $h_1$ is trivial, that is, if it is a constant,
we may set $h_1=0$ without loss of generality.
This means we have $F_{(5)}=0$ and $h=h_0(x)$. It then follows
from (\ref{D3D7-2:D3-D7 brane Einstein equation-ij2:Eq}) that
$\lap_{\Xsp}h_0=0$. 
If we assume $q_{\mu\nu}=\eta_{\mu\nu}$, $s_{ab}=\delta_{ab}$, 
the solution for $h_0$ is given by
\Eq{
h_0(x)=c_{\mu}x^{\mu}+d,
}
in terms of the four-dimensional Minkowski coordinates $x^{\mu}$, 
where $c_{\mu}$ and $d$ are constants.

Equation~(\ref{D3D7-2:warp factor of D3-D7 brane solution:Eq}) 
implies that the function $h_0$ is in the same form as 
in the case of a single D3-brane solution with $\lambda=0$
as seen from \eqref{eq:sec2:warp factor of p-brane solution c=0}.
Thus, we find that the metric for the D3-D7 brane system
(\ref{D3D7-2:metric of solution:Eq}) is similar to that of 
the D3-brane system. 
Apart from the restriction that $\lambda=0$, the difference from
the D3-brane metric is in the six-dimensional metric $u_{ij}$,
which is affected by the existence of a nontrivial scalar 
field configuration, which describes a D7 brane.

The ten-dimensional metric 
(\ref{D3D7-2:metric of solution:Eq}) exists for $h>0$ and 
has curvature singularities at $h=0$, $r=0$, and $r=r_{\rm c}$. 
If the function $h_1$ is negligible and $h_0(t)\propto t$, 
the scale factor of the four-dimensional universe is 
$\propto\tau^{-2/3}$, while the scale factor of the six-dimensional 
internal space is $\propto\tau^{2/3}$, where 
$\tau\propto t^{3/4}$ is the cosmic time in four dimensions.
Thus, as in the case of the D4-D8 brane system,
the D3-D7 solution also behaves as a Kaluza-Klein type cosmological
solution in the asymptotic future. On the other hand, the analysis
of the solution near $t=0$ is not so easy compared to the case of
the D4-D8 brane system. Therefore, we postpone it as a future issue.

We can construct the D3-D7-brane solution on the basis of the 
discussion in Sec.~\ref{sec:Dynamical p-brane solution}.
Now we assume the ten-dimensional metric of the form
\Eqr{
ds^2&=&h_3^{-1/2}(x, y)q_{\mu\nu}(\Xsp)dx^{\mu}dx^{\nu}
+h_3^{1/2}(x, y)\gamma_{ab}(\Zsp)dz^{a}dz^{b}\nn\\
&&+h_3^{1/2}(x, y)h_7(x, y)u_{ij}(\Ysp)dy^idy^j,
\label{D3D7-3:metric:Eq}
}
where $q_{\mu\nu}(\Xsp)$ denotes a four-dimensional metric, which
depends only on the four-dimensional coordinates $x^{\mu}$, 
$\gamma_{ab}(\Zsp)$ denotes a four-dimensional metric, which
depends only on the four-dimensional coordinates $z^{a}$, 
and $u_{ij}(\Ysp)$ is the two-dimensional metric, which
depends only on the two-dimensional coordinates $y^i$. 
We also assume that the dilaton $\phi$ and the gauge 
fields are given by
\Eqr{
F_{(5)}&=&(1\pm\ast)d\left(h_5^{-1}\right)\wedge \Omega(\Xsp),
   \label{D3D7-3:F5:Eq}\\
F_{(1)}&=&e^{-2\phi}\ast F_{(9)}=e^{-2\phi}\ast\left[
d\left(h_7^{-1}\right)\wedge \Omega(\Xsp)\wedge \Omega(\Zsp)\right],
   \label{D3D7-3:F9:Eq}\\
e^{\phi}&=&h_7^{-1},
   \label{D3D7-3:dilaton:Eq}
}
where $\Omega(\Xsp)$ and $\Omega(\Zsp)$ are given by 
\Eq{
\Omega(\Xsp)=\sqrt{-q}dx^0\wedge\cdots\wedge dx^3,~~~~
\Omega(\Zsp)=\sqrt{\gamma}dz^4\wedge\cdots\wedge dz^7.
 }

The dilaton and gauge fields are assumed to have
the same form as the single $p$-brane solutions 
in Sec.~\ref{sec:Dynamical p-brane solution}.
The assumptions on the ten-dimensional metric and fields 
correspond to the following brane configuration:
\begin{center}
\begin{tabular}{|c|c|c|c|c|c|c|c|c|c|c|}
\hline
&0&1&2&3&4&5&6&7&8&9\\
\hline
D3 & $\circ$ & $\circ$ & $\circ$ & $\circ$ &&&&&& \\
\hline
D7 & $\circ$ & $\circ$ & $\circ$ & $\circ$ & $\circ$ & $\circ$ 
& $\circ$ & $\circ$ & & \\ 
\hline
\end{tabular}
\end{center}
This is the same as the previous D3-D7-brane configuration 
(\ref{D3D7-2:metric of solution:Eq}).
We first consider the Einstein 
Eqs.~(\ref{D3D7-2:Einstein equation:Eq}). Substituting the metric 
(\ref{D3D7-3:metric:Eq}) into the ten-dimensional Einstein equations,
we obtain
\Eqr{
&&R_{\mu\nu}(\Xsp)-h_3^{-1}D_{\mu}D_{\nu} h_3 -h_7^{-1}D_{\mu}D_{\nu} h_7 
-\frac{1}{2}\left(\pd_{\mu}\ln h_3\pd_{\nu}\ln h_7
+\pd_{\mu}\ln h_7\pd_{\nu}\ln h_3\right)\nn\\
&&~~~~+\frac{1}{4}h_3^{-1} 
q_{\mu\nu}\left[\triangle_{\Xsp}h_3 + \left(h_3h_7\right)^{-1}
\left(\triangle_{\Ysp} h_3
+h_3q^{\rho\sigma}\pd_{\rho}h_3\pd_{\sigma}h_7\right)\right]
=0,
 \label{D3D7-3:Einstein mu:Eq}\\
&&R_{ab}(\Zsp)
-\frac{1}{4}\gamma_{ab}\left[\triangle_{\Xsp}h_3 + \left(h_3h_7\right)^{-1}
\left(\triangle_{\Ysp} h_3
+h_3q^{\rho\sigma}\pd_{\rho}h_3\pd_{\sigma}h_7\right)\right]=0,
 \label{D3D7-3:Einstein ab:Eq}\\
&&R_{ij}(\Ysp)-\frac{1}{4} u_{ij}\left(h_7\triangle_{\Xsp} h_3
 +h_3^{-1}\triangle_{\Ysp}h_3 \right)
 -\frac{1}{2} u_{ij}\left(h_3\triangle_{\Xsp} h_7
 +h_7^{-1}\triangle_{\Ysp}h_7 \right)\nn\\
&&~~~~-\frac{3}{4}u_{ij}q^{\rho\sigma}\pd_{\rho}h_3\pd_{\sigma}h_7=0,
 \label{D3D7-3:Einstein ij:Eq}\\
&&h_3^{-1}\pd_{\mu}\pd_i h_3+h_7^{-1}\pd_{\mu}\pd_i h_7=0,
 \label{D3D7-3:Einstein mi:Eq}
}
where $D_{\mu}$ is the covariant derivative with respect to 
the metric $q_{\mu\nu}$, 
$\triangle_{\Xsp}$ and $\triangle_{\Ysp}$ are 
the Laplace operators on the space of 
${\rm \Xsp}$ and the space ${\rm \Ysp}$, respectively, and 
$R_{\mu\nu}(\Xsp)$, $R_{ab}(\Zsp)$, and $R_{ij}(\Ysp)$ are 
the Ricci tensors
of the metrics $q_{\mu\nu}$, $\gamma_{ab}$, and $u_{ij}$, respectively.
From \eqref{D3D7-3:Einstein mi:Eq}, we have
\Eq{
h_3(x, y)= H_0(x)+H_1(y),~~~~h_7(x, y)= K_0(x)+K_1(y).
  \label{D3D7-3:h:Eq}
}
Using (\ref{D3D7-3:h:Eq}),
Eqs.~(\ref{D3D7-3:Einstein mu:Eq}), (\ref{D3D7-3:Einstein ab:Eq}) 
and (\ref{D3D7-3:Einstein ij:Eq}) 
are rewritten as
\Eqr{
&&R_{\mu\nu}(\Xsp)-h_3^{-1}D_{\mu}D_{\nu} H_0 -h_7^{-1}D_{\mu}D_{\nu} K_0 
-\frac{1}{2}\left(h_3h_7\right)^{-1}
\left(\pd_{\mu}H_0\pd_{\nu}K_0
+\pd_{\mu}K_0\pd_{\nu}H_0\right)\nn\\
&&~~~~+\frac{1}{4}h_3^{-1} 
q_{\mu\nu}\left[\triangle_{\Xsp}H_0 + \left(h_3h_7\right)^{-1}
\left(\triangle_{\Ysp} H_1
+h_3q^{\rho\sigma}\pd_{\rho}H_0\pd_{\sigma}K_0\right)\right]=0,
 \label{D3D7-3:Einstein mu2:Eq}\\
&&R_{ab}(\Zsp)
-\frac{1}{4}\gamma_{ab}\left[\triangle_{\Xsp}H_0 + \left(h_3h_7\right)^{-1}
\left(\triangle_{\Ysp} H_1
+h_3q^{\rho\sigma}\pd_{\rho}H_0\pd_{\sigma}K_0\right)\right]=0,
 \label{D3D7-3:Einstein ab2:Eq}\\
&&R_{ij}(\Ysp)-\frac{1}{4} u_{ij}\left(h_7\triangle_{\Xsp} H_0
 +h_3^{-1}\triangle_{\Ysp}H_1 \right)
 -\frac{1}{2} u_{ij}\left(h_3\triangle_{\Xsp} K_0
 +h_7^{-1}\triangle_{\Ysp}K_1 \right)\nn\\
&&~~~~-\frac{3}{4}u_{ij}q^{\rho\sigma}\pd_{\rho}H_0\pd_{\sigma}K_0=0.
 \label{D3D7-3:Einstein ij2:Eq}
   }

We can simplify Eqs.~(\ref{D3D7-3:Einstein mu2:Eq})
-(\ref{D3D7-3:Einstein ij2:Eq}) 
in terms of the gauge and dilaton equations.
Under the assumption (\ref{D3D7-3:F5:Eq}), 
we find
\Eq{
dF_{(5)}=\mp \triangle_{\Ysp}H_1 \,\Omega(\Zsp)\wedge 
dy^i\wedge \ast_{\Ysp} dy^i
=0,
}
where we have used (\ref{D3D7-3:h:Eq}), 
and $\ast_{\Ysp}$ denotes the Hodge dual operator on $\Ysp$.
Thus, we obtain the equation,
\Eq{
\lap_{\Ysp}H_1=0.
 \label{D3D7-3:H1:Eq}
}
The scalar field 
Eq.~(\ref{D3D7-2:scalar field equation:Eq}) 
can be decomposed as follows
\Eqr{
&&d(e^{2\phi}\ast F_{(1)})=0,
   \label{D3D7-3:F1 eq:Eq}\\
&&\lap\phi=e^{2\phi}F_{(1)}^2.
   \label{D3D7-3:dilaton eq:Eq}
}
Under \eqref{D3D7-3:F9:Eq}, the field 
Eq.~(\ref{D3D7-3:F1 eq:Eq}) 
becomes the Bianchi identity for $F_{(9)}$. 
Then, the field equation for $F_{(1)}$ 
is automatically satisfied. On the other hand, 
from the Bianchi identity for $F_{(1)}$, we have
\Eqr{
d(e^{-2\phi}\ast F_{(9)})=
-\triangle_{\Ysp}K_1 \,dy^i\wedge \ast_{\Ysp} dy^i=0.
}
Then, the Bianchi identity for $F_{(1)}$ leads to
\Eq{
\lap_{\Ysp}K_1=0.
 \label{D3D7-3:K1:Eq}
}
Next we consider the dilaton 
Eq.~(\ref{D3D7-3:dilaton eq:Eq}).
Substituting Eqs.~(\ref{D3D7-3:F5:Eq}) - (\ref{D3D7-3:dilaton:Eq}), 
and the metric (\ref{D3D7-3:metric:Eq}) 
into the equation of motion for the dilaton  
(\ref{D3D7-3:dilaton eq:Eq}), we find 
\Eq{
h_3^{-1/2}h_7^{-1}\left(q^{\mu\nu}\pd_{\mu}H_0\pd_{\nu}K_0
+h_3\lap_{\Xsp}H_0\right)=0,
   \label{D3D7-3:dilaton eq2:Eq}
}
where we used Eqs.~(\ref{D3D7-3:h:Eq}), (\ref{D3D7-3:K1:Eq}).
Then, in order to satisfy field 
Eqs.~(\ref{D3D7-3:Einstein mu2:Eq}) - 
(\ref{D3D7-3:Einstein ij2:Eq}), (\ref{D3D7-3:H1:Eq}), 
(\ref{D3D7-3:K1:Eq}), (\ref{D3D7-3:dilaton eq2:Eq}), 
we can choose either
\Eqr{
&&R_{\mu\nu}(\Xsp)=0, ~~~R_{ab}(\Zsp)=0,~~~R_{ij}(\Ysp)=0,
   \label{D3D7-3:Ricci1:Eq}\\
&&D_{\mu}D_{\nu}H_0=0,~~~K_0={\rm constant},
   \label{D3D7-3:warp1:Eq} 
 }
or 
\Eqr{
&&R_{\mu\nu}(\Xsp)=0, ~~~R_{ab}(\Zsp)=0,~~~R_{ij}(\Ysp)=0,
   \label{D3D7-3:Ricci2:Eq}\\
&&D_{\mu}D_{\nu}K_0=0,~~~H_0={\rm constant}.
   \label{D3D7-3:warp2:Eq} 
 }
We find that the Einstein equations cannot be satisfied if 
both $h_3$ and $h_7$ depend on the coordinate $x^{\mu}$.

If we assume $\pd_{\mu}h_7=0$, $q_{\mu\nu}=\eta_{\mu\nu}$, 
$\gamma_{ab}=\delta_{ab}$, $u_{ij}=\delta_{ij}$,  we have
\Eqr{
&&h_3(x, y)=H_0(x)+H_1(y);~~H_0(x)=c_{\mu}x^{\mu}+d,~~~
H_1(y)=\sum_l M_l\ln|\vec{y}-\vec{y}_l|,\\
&&h_7(y)=\sum_l L_l\ln|\vec{y}-\vec{y}_l|,
}
where $c_{\mu}$, $d$, $M_l$, $L_l$ and $\vec{y}_l$ are constants. 
For $\pd_{\mu}h_3=0$, $q_{\mu\nu}=\eta_{\mu\nu}$, 
$\gamma_{ab}=\delta_{ab}$, $u_{ij}=\delta_{ij}$, 
we have
\Eqr{
&&h_7(x, y)=K_0(x)+K_1(y);~~K_0(x)=c_{\mu}x^{\mu}+d,~~~
K_1(y)=\sum_l M_l\ln|\vec{y}-\vec{y}_l|,\\
&&h_3(y)=\sum_l L_l\ln|\vec{y}-\vec{y}_l|, 
}
where $c_{\mu}$, $d$, $M_l$, $L_l$, and $\vec{y}_l$ are constants.

There are curvature singularities at $\vec{y}=\vec{y}_l$ 
as well as at $h_3=0$ or $h_7=0$ in the ten-dimensional metric 
(\ref{D3D7-3:metric:Eq}). 
In the case of $\pd_{\mu}h_7=0$, the function $H_0$ depends on a
linear function of the four-dimensional coordinates $x^{\mu}$. 
The scale factor of the four-dimensional 
universe has the same form as the previous D3-D7-brane solution 
(\ref{D3D7-2:metric of solution:Eq}). For $\pd_{\mu}h_3=0$, 
the function $K_0$ is proportional to a linear function of the 
four-dimensional coordinates $x^{\mu}$.
The scale factor $a(t)$ of the two-dimensional internal space $\Ysp$ 
is given by $a(t)\propto t$ if $K_1=0$ and $K_0(t)\propto t$. 
 
The D3-D7-brane metric (\ref{D3D7-3:metric:Eq}) 
is not the same as (\ref{D3D7-2:metric of solution:Eq}) because 
it has the dilaton and gauge field, 
which depend on the four-dimensional coordinate $x^{\mu}$. 
In particular, the two-dimensional internal space $\Ysp$ depends
on time even if $\pd_{\mu}h_3=0$.

\section{Conclusion}
  \label{sec:Discussions}
In this paper, we investigated dynamical solutions 
of higher-dimensional supergravity models. 
We found a class of time-dependent solutions for 
intersecting D4-D8- and D3-D7-brane systems. 
These solutions were obtained by replacing a
constant $A$ in the warp factor $h=A+h_1(y)$ of
a supersymmetric solution by a function $h_0(x)$ of the 
 coordinates $x^\mu$~\cite{Kodama:2005fz},
where the coordinates $y^i$ would describe the 
internal space and $x^\mu$ would describe our Universe.
In the D4-D8 brane solution, the geometry was found to 
approach a warped static ${\rm AdS}_6 \times {\rm S}^4$
in a certain region of the spacetime.

In particular, in the case of the D4-D8 system,
we found an interesting solution
which is warped and static as $\tau\to0$ but
approaches a Kasner-type solution as $\tau\to\infty$,
where $\tau$ is the cosmic time.
Although the solution itself is by no means realistic,
its interesting behavior suggests a possibility that 
the Universe was originally in a static state of warped 
compactification and began to evolve toward a Universe
with a Kaluza-Klein compactified internal space.

Conventionally one would expect an effective theory description
in lower dimensions to be valid at low energies.
However, as clearly the case of the cosmological solution
mentioned above, the solutions we found have 
the property that they are genuinely $D$ dimensional in the
sense that one can never neglect the dependence on $y^i$,
say of $h$. 
Thus, our result indicates that we have to be careful 
when we use a four-dimensional effective theory to analyze 
the moduli stabilization problem and the cosmological 
problems in the framework of warped compactification of 
supergravity or M theory.

\section*{Acknowledgments}

This work was supported by the CNRS-JSPS Joint Research Program
and by JSPS Grant-in-Aid for Scientific Research (B), 
under Contract Nos.~17340075 and 
(A) 18204024, and also by JSPS Grant-in-Aid for Creative 
Scientific Research, under Contract No.~19GS0219. 
KU is supported by the Kwansei Gakuin University, 
Grant-in-Aid for Young Scientists (B) of JSPS Research, 
under Contract No.~20740147, 
and he is grateful to Shigeki Sugimoto for valuable discussions.




\section*{References}
\begin{thebibliography}{10}

\bibitem{Kodama:2005cz}
  H.~Kodama and K.~Uzawa,
  ``Comments on the four-dimensional effective theory for warped
  compactification,''
  JHEP {\bf 0603} (2006) 053
  [arXiv:hep-th/0512104].
  
\bibitem{Gibbons:2005rt}
  G.~W.~Gibbons, H.~Lu and C.~N.~Pope,
  ``Brane worlds in collision,''
  Phys.\ Rev.\ Lett.\  {\bf 94} (2005) 131602
  [arXiv:hep-th/0501117].

\bibitem{Duff:1986hr}
  M.~J.~Duff, B.~E.~W.~Nilsson and C.~N.~Pope,
  ``Kaluza-Klein supergravity,''
  Phys.\ Rept.\  {\bf 130} (1986) 1.

\bibitem{Ohta:2006sw}
  N.~Ohta and K.~L.~Panigrahi,
  ``Supersymmetric intersecting branes in time-dependent backgrounds,''
  Phys.\ Rev.\  D {\bf 74} (2006) 126003
  [arXiv:hep-th/0610015].
  
\bibitem{Gueven:1992hh}
  R.~G\"{u}ven,
  ``Black p-brane solutions of D = 11 supergravity theory,''
  Phys.\ Lett.\  B {\bf 276} (1992) 49.

\bibitem{Polchinski:1995df}
  J.~Polchinski and E.~Witten,
  ``Evidence for heterotic - type I string duality,''
  Nucl.\ Phys.\  B {\bf 460} (1996) 525
  [arXiv:hep-th/9510169].
  
\bibitem{Youm:1999zs}
  D.~Youm,
  ``Localized intersecting BPS branes,''
  arXiv:hep-th/9902208.

\bibitem{Ohta:1997gw}
  N.~Ohta,
  ``Intersection rules for non-extreme p-branes,''
  Phys.\ Lett.\  B {\bf 403} (1997) 218
  [arXiv:hep-th/9702164].

\bibitem{Nastase:2003dd}
  H.~Nastase,
  ``On Dp-Dp+4 systems, QCD dual and phenomenology,''
  arXiv:hep-th/0305069.

\bibitem{Brandhuber:1999np}
  A.~Brandhuber and Y.~Oz,
  ``The D4-D8 brane system and five dimensional fixed points,''
  Phys.\ Lett.\  B {\bf 460} (1999) 307
  [arXiv:hep-th/9905148].

\bibitem{Romans:1985tz}
  L.~J.~Romans,
  ``Massive N=2a supergravity in ten-dimensions,''
  Phys.\ Lett.\  B {\bf 169} (1986) 374.

\bibitem{Romans:1985tw}
  L.~J.~Romans,
  ``The F(4) gauged supergravity in six-dimensions,''
  Nucl.\ Phys.\  B {\bf 269} (1986) 691.

\bibitem{Cvetic:1999un}
  M.~Cvetic, H.~Lu and C.~N.~Pope,
  ``Gauged six-dimensional supergravity from massive type IIA,''
  Phys.\ Rev.\ Lett.\  {\bf 83} (1999) 5226
  [arXiv:hep-th/9906221].  

\bibitem{Aharony:1998xz}
  O.~Aharony, A.~Fayyazuddin and J.~M.~Maldacena,
  ``The large N limit of N = 2,1 field theories from three-branes in
  F-theory,''
  JHEP {\bf 9807} (1998) 013
  [arXiv:hep-th/9806159].

\bibitem{Grana:2001xn}
  M.~Grana and J.~Polchinski,
  ``Gauge / gravity duals with holomorphic dilaton,''
  Phys.\ Rev.\  D {\bf 65} (2002) 126005
  [arXiv:hep-th/0106014].

\bibitem{Burrington:2004id}
  B.~A.~Burrington, J.~T.~Liu, L.~A.~Pando Zayas and D.~Vaman,
  ``Holographic duals of flavored N = 1 super Yang-Mills: Beyond the probe
  approximation,''
  JHEP {\bf 0502} (2005) 022
  [arXiv:hep-th/0406207].

\bibitem{Kirsch:2005uy}
  I.~Kirsch and D.~Vaman,
  ``The D3/D7 background and flavor dependence of Regge trajectories,''
  Phys.\ Rev.\  D {\bf 72} (2005) 026007
  [arXiv:hep-th/0505164].

\bibitem{Lu:1995cs}
  H.~Lu, C.~N.~Pope, E.~Sezgin and K.~S.~Stelle,
  ``Stainless super p-branes,''
  Nucl.\ Phys.\  B {\bf 456} (1995) 669
  [arXiv:hep-th/9508042].

\bibitem{Stelle:1998xg}
  K.~S.~Stelle,
  ``BPS branes in supergravity,''
  arXiv:hep-th/9803116.
  
\bibitem{Argurio:1998cp}
  R.~Argurio,
  ``Brane physics in M-theory,''
  arXiv:hep-th/9807171.

\bibitem{Kodama:2005fz}
  H.~Kodama and K.~Uzawa,
  ``Moduli instability in warped compactifications of the type IIB
  supergravity,''
  JHEP {\bf 0507} (2005) 061
  [arXiv:hep-th/0504193].

\bibitem{Bergshoeff:1996ui}
  E.~Bergshoeff, M.~de Roo, M.~B.~Green, G.~Papadopoulos and P.~K.~Townsend,
  ``Duality of Type II 7-branes and 8-branes,''
  Nucl.\ Phys.\  B {\bf 470} (1996) 113
  [arXiv:hep-th/9601150].
  
\bibitem{Bergshoeff:1997ak}
  E.~Bergshoeff, Y.~Lozano and T.~Ortin,
  ``Massive branes,''
  Nucl.\ Phys.\  B {\bf 518} (1998) 363
  [arXiv:hep-th/9712115].

\bibitem{Imamura:2001cr}
  Y.~Imamura,
  ``1/4 BPS solutions in massive IIA supergravity,''
  Prog.\ Theor.\ Phys.\  {\bf 106} (2001) 653
  [arXiv:hep-th/0105263].
    
\bibitem{Cvetic:1999xx}
  M.~Cvetic, S.~S.~Gubser, H.~Lu and C.~N.~Pope,
  ``Symmetric potentials of gauged supergravities in diverse dimensions and
  Coulomb branch of gauge theories,''
  Phys.\ Rev.\  D {\bf 62} (2000) 086003
  [arXiv:hep-th/9909121].

\bibitem{Behrndt:1999mk}
  K.~Behrndt, E.~Bergshoeff, R.~Halbersma and J.~P.~van der Schaar,
  ``On domain-wall/QFT dualities in various dimensions,''
  Class.\ Quant.\ Grav.\  {\bf 16} (1999) 3517
  [arXiv:hep-th/9907006].

\bibitem{Nunez:2001pt}
  C.~Nunez, I.~Y.~Park, M.~Schvellinger and T.~A.~Tran,
  ``Supergravity duals of gauge theories from F(4) gauged supergravity in  six
  dimensions,''
  JHEP {\bf 0104} (2001) 025
  [arXiv:hep-th/0103080].

\bibitem{Schwarz:1983qr}
  J.~H.~Schwarz,
  ``Covariant field equations of chiral N=2 D=10 supergravity,''
  Nucl.\ Phys.\  B {\bf 226} (1983) 269.

\bibitem{Greene:1989ya}
  B.~R.~Greene, A.~D.~Shapere, C.~Vafa and S.~T.~Yau,
  ``Stringy cosmic strings and noncompact Calabi-Yau manifolds,''
  Nucl.\ Phys.\  B {\bf 337} (1990) 1.


\end{thebibliography}
\end{document}